\documentclass[prd,twocolumn]{revtex4}
\usepackage{graphicx}
\usepackage{amssymb}
\usepackage{color}
\usepackage{enumerate}
\usepackage{amsmath}

\newcommand{\be}{\begin{equation}}
\newcommand{\ee}{\end{equation}}
\newcommand{\bea}{\begin{eqnarray}}
\newcommand{\eea}{\end{eqnarray}}

\begin{document}

	\title{Note on complex metrics, complex time and periodic universes}

	\author{Fabio Briscese}\email{fabio.briscese@math.unipd.it, briscese.phys@gmail.com}
	
	\affiliation{Department of Mathematics “Tullio Levi-Civita”,
		University of Padova, via Trieste, 63 - 35121, Italy}
		\affiliation{ Istituto Nazionale di Alta Matematica Francesco		Severi, Gruppo Nazionale di Fisica Matematica, Citt\`{a} Universitaria, P.le A. Moro 5, 00185 Rome, Italy.}

	\begin{abstract}
	
		Motivated, on the one hand, by recent results on isochronous dynamical systems, and on the other, by quantum gravity  applications of complex metrics, we show that, if such an enlarged class of metrics is considered, one can easily obtain periodic or bouncing complex solutions of Einstein's equations. It is found that, for any given  solution $g_{\mu\nu}$ of the Einstein's equations, by means of a complex  periodic change of time, one can construct infinitely many  periodic or bouncing complex solutions $\hat g_{\mu\nu}$ that are  physically indistinguishable from  $g_{\mu\nu}$ over an arbitrarily long time interval. 	These results, which are based on the use of complex diffeomorphisms, 	point out an unacceptable arbitrariness in the theory. As we will show, a condition on the class of physically meaningful complex metrics proposed in \cite{Kontsevich} and discussed in \cite{Witten} solves this problem, restricting the family of admissible
		complex diffeomorphisms. We conclude by arguing that this condition can be viewed as a  quantum-gravity generalization of the equivalence principle to complex spacetimes.

	\end{abstract}
	
	\maketitle

The study of isochronous dynamical systems (namely, dynamical systems that admit only periodic solutions with the same period) has taken a significant step forward when it has been shown that, given a nonrelativistic autonomous dynamical system $D$, other autonomous dynamical systems
$\hat{D}$ can be manufactured, featuring two additional arbitrary positive parameters $T$ and $\hat{T}$ with $T>\hat{T}$ having the following two
properties: (i) For the same variables of the original dynamical system $D$
the new dynamical system $\hat{D}$ yields, over the time interval $\hat{T}$, hence for an \textit{arbitrarily long} time, a dynamical
evolution which mimics \textit{arbitrarily closely} that yielded
by the original system $D$; up to corrections of order
$t/\hat{T}$, or possibly even \textit{identically}. (ii) The system $\tilde{D}$ is \textit{isochronous}; see \cite{calogero1,calogero2,calogero3,calogero4,calogero6,calogero7,calogero8,calogero book} for review. Among other things, this result implies that isochronous systems are not rare. Moreover, it also entails the unpleasant fact that, for any dynamical system it is possible to build infinite dynamical systems which cannot be distinguished experimentally by means of any experiment that lasts a finite time.

As this result is valid nonrelativistic dynamical systems, including the case of  N-body systems with an arbitrarily large number of bodies, the question whether it could be extended to the cosmological context arises quite naturally.
This issue has been studied in a series of papers \cite{isochronous cosmologies,isochronous spacetimes,isochronous Newtonian limit} where it has been shown that,  for any given real solution $g_{\mu\nu}$ of  Einstein's equations, one can construct real periodic (with an arbitrarily long period $T$) solutions $\hat g_{\mu\nu}$ that are degenerate on a countable infinite set of times $t_n$ and diffeomorphic to $g_{\mu\nu}$ elsewhere, and are thus physically indistinguishable from $g_{\mu\nu}$  at any time $t\neq t_n$, whose physical interpretation can be given in terms of a version of general relativity in which the equivalence principle is lost.

The trick used for obtaining the nonrelativistic isochronous systems $\hat{D}$ is based on the introduction of a periodic change of time $t \rightarrow \tau(t)$ with $\tau(t)$ periodic with arbitrary period $T$, such that $\tau(t)\simeq t$  for any $t$ in an interval $[t_0-\hat{T}/2,t_0+\hat{T}/2]$. 
If one tries to extend this trick to general relativity, one immediately realizes that a real periodic change of time $\tau(t)$  leads to  periodic solutions of  Einstein's equations that are degenerate at the numerable inversion times $t_n$, where $\tau^\prime(t_n)=0$. 
However, the equivalence principle implies that physically acceptable spacetimes must be locally Minkowskian. Indeed, the method used for manufacturing the isochronous dynamical systems $\hat{D}$ does not work for general relativity, as the equivalence principle forbids periodic changes of time, avoiding  the unpleasant implications that occur for nonrelativistic systems.

\begin{figure}	\begin{center}
		\includegraphics[height=4cm]{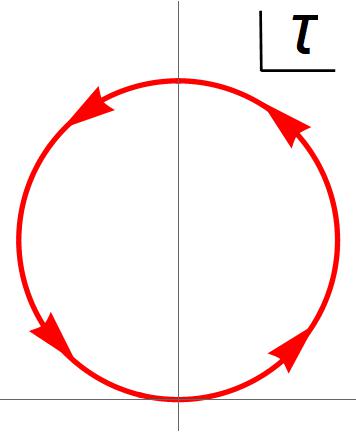}	
		\caption{We plot the periodic change of time $\tau(t)=T\left(i+\exp\left[i(t/T-\pi/2)\right]\right)$. One has $\tau^\prime(t)\neq0$ and $\tau\simeq t$ for $|t|\ll T$. }\label{fig12}
	\end{center}
\end{figure}

In this paper we point out that, if the class of admissible spacetimes is enlarged to include complex metrics, one has to face the same issues encountered in the case of nonrelativistic systems. In fact, one could consider a complex and periodic change of time $t\rightarrow\tau(t)$  with $\tau^\prime(t)\neq 0$. For instance, considering the function

\be
\tau(t)= T\left(i+\exp\left[i(t/T-\pi/2)\right]\right)
\ee
plotted  in Fig. \ref{fig12}, and a real Friedmann-Robertson-Walker (FRW) spacetime

\be\label{FRW1}
ds^2 = dt^2 + a(t)^2 \, d\vec{x}^2 \, ,
\ee
one obtains a metric
\be\label{flat metric complex 2}
ds^2 = \tau^\prime(t)^2 dt^2 + a(\tau(t))^2 \, d\vec{x}^2 \, .
\ee
The metric (\ref{flat metric complex 2}) is non degenerate because $\tau^\prime(t)\neq 0$ at any time, and it is diffeomorphic to (\ref{FRW1}), indeed all the curvature invariants of (\ref{flat metric complex 2}) can be calculated from those of the real metric (\ref{FRW1}), provided that $\tau(t) \in \mathcal{C}^3$. 
Moreover, at any time  $|t|\ll T$,   (\ref{FRW1}) and  (\ref{flat metric complex 2}) will be the same up to corrections $\sim O(t/T)$. This implies that, since $T$ is arbitrary, the two metrics (\ref{FRW1}) and  (\ref{flat metric complex 2}) will be physically indistinguishable over an arbitrarily long time interval, albeit being globally inequivalent, since (\ref{flat metric complex 2}) is periodic (we will discuss this point and the equivalence between diffeomorphic complex metrics in what follows).

This result can be generalized to state that, by means of a complex periodic change of time, for any solution $g_{\mu\nu}$ of  Einstein's equations one can easily obtain nondegenerate but complex and periodic metrics $\hat{g}_{\mu\nu}$ that are indistinguishable from $g_{\mu\nu}$  over an arbitrarily long time interval. What is more, we will show that one can also  build bouncing solutions corresponding to an arbitrary inversion of the arrow of time, as well as (periodic or bouncing) solutions that avoid the big-bang cosmological singularity.  We should probably emphasize that these results do not depend on the fact that the starting metric $g_{\mu \nu}$ is real.  The key observation is that, as long as complex diffeomorphisms are allowed, one can write down periodic changes of coordinates that are locally invertible everywhere, leading to nondegenerate periodic metrics $\hat g_{\mu \nu}$ that solve  Einstein's equations everywhere (except at most at the singularities of $g_{\mu \nu}$). The same conclusions are valid also for bouncing spacetimes, i.e. for time inversion.

Of course, the implications of these findings are problematic, as it would be impossible to make an experiment lasting a finite amount of time which is able to discern between a periodic and nonperiodic universe, or between a universe facing a big-bang singularity and another which is singularity free. In the case of real spacetimes, the equivalence principle prevents the occurrence of these issues, since it forbids periodic diffeomorphisms. However, as long as one considers complex metrics, the equivalence principle does not exclude complex periodic changes of time, and we do not have any other criterion capable of doing so in the context of classical gravitation.

The interest in complex spacetimes, which also motivates this paper, stays in the fact that they are widely used in the framework of quantum field theory on curved spacetimes and in quantum gravity.
One of the first attempts to make sense of complex metrics dates back to the late 1970s \cite{Gibbson}, when it has been noted that the Kerr metric, unlike the Schwarzschild spacetime, becomes complex after a Wick rotation. It has been also observed that a complex metric could be useful to provide a positive-defined quantum-gravity action, in order to give convergent path integrals \cite{Gibbson 2}. Without aiming to be exhaustive, we just mention that complex metrics have been considered in the study of the Hartle-Hawking no-boundary proposal for the wave function of the universe \cite{Haliwell,Hartle} and of the topology-changing processes \cite{Louko}. For more recent applications  of complex spacetimes  for the definition of the Euclidean path integral of general relativity see \cite{Kontsevich,Witten, a1,a2,a3,a4,a5,a6,a7,a8,a9,a10,a11,a12,a13,a14,a15,a16,a17,a18,a19,a20}.

Very recently, in \cite{Kontsevich}  a criterion for defining the allowed class of complex metrics  has been proposed. Such criterion is based on the request that a quantum field theory can be consistently defined in the corresponding  spacetime. This proposal has  been discussed further in \cite{Witten} in the context of quantum gravity. Namely, a physically acceptable complex metric $g_{\mu\nu}$ must be such that, writing $g_{\mu\nu}$ in a orthogonal basis as $g_{\mu\nu}=\delta_{\mu\nu} \, \lambda_\mu$, the condition

\begin{equation}\label{criterion}
\lambda_\mu\neq 0;\qquad \sum_{\mu=1}^{4} |Arg(\lambda_\mu)|<\pi \, ,
\end{equation}
is satisfied, where $Arg(z)\in[-\pi,\pi]$ is the argument of $z\in\mathcal{C}$. We direct the reader to \cite{Kontsevich,Witten} for the derivation of this condition, since here we are only interested to its implications on the construction of isochronous spacetimes.

As we will show below,  the condition (\ref{criterion}) plays a role analogous to that played by the equivalence principle in the case of general relativity: it forbids the use of periodic complex changes of time (and complex time inversion), preventing the occurrence of the issues related to the construction of isochronous and bouncing metrics. Indeed, this criterion can be viewed as a generalization of the equivalence principle for complex spacetimes.

In what follows we will first give an illustration of the method used for finding nonrelativistic isochronous systems based non N-body Hamiltonians. Then we will discuss the generalization of this method to spacetime metrics and the issues related to the construction of isochronous and bouncing solutions of Einstein's equations. Finally, we will discuss the implications of the condition (\ref{criterion}) and show that it forbids the class of complex and periodic changes of times used for finding the isochronous metrics. This motivates our conclusion that (\ref{criterion}) can be viewed as a quantum-gravity generalization of the equivalence principle.

As stated before, the method used for finding the isochronous systems $\hat{D}$ from an autonomous dynamical system $D$ is based on the introduction of a periodic change of time $t \rightarrow \tau(t)$ with $\tau(t)$ periodic with an arbitrary period $T$, such that $\tau(t)\simeq t$  for  $t\in [t_0-\hat{T}/2,t_0+\hat{T}/2]$. For instance, one can start from an N-body Hamiltonian

\begin{equation}
H\left( \underline{p},\underline{q}\right) =\frac{1}{2}\sum_{n=1}^{3N}\left(
p_{n}^{2}\right) +V\left( \underline{q}\right)  \label{H}
\end{equation}%
with a translation-invariant potential $V\left( \underline{q}%
+a\right) =V\left( \underline{q}\right) $ and solutions $\underline{q}(t)$, $\underline{p}(t)$. 
Then, one switches to the center-of-mass reference system defining the coordinates $x_{n}=q_{n}-Q\left( \underline{q}
\right) ,$ $y_{n}=p_{n}-P\left( \underline{p}\right) /N$ , where $Q\left( \underline{q}\right) =
\sum_{n=1}^{N}q_{n}/N$ and $P\left( \underline{p}\right)
=\sum_{n=1}^{N}p_{n}$ are the coordinates of the center-of-mass and the total momentum, so that

\begin{equation}
H\left( \underline{p},\underline{q}\right) =\frac{\left[ P\left( \underline{p%
	}\right) \right] ^{2}}{2N}+h\left( \underline{y},\underline{x}\right) \, ,  
\end{equation}
where
\begin{equation}
h\left( \underline{y},\underline{x}\right) =\frac{1}{2}\sum_{n=1}^{N}\left(
y_{n}^{2}\right) +V\left( \underline{x}\right) ~,  
\end{equation}
and $V\left( \underline{x}\right) =V\left( \underline{q}\right) $ thanks
to the assumed translation invariance of $V\left( \underline{q}\right) $. It can be shown \cite{calogero1,calogero2,calogero3,calogero4,calogero6,calogero7,calogero8,calogero book}  that the following 
\textit{isochronous} Hamiltonian $\tilde{H}\left( \underline{p},\underline{q}%
;T\right) $ 

\begin{equation}
\hat{H}\left( \underline{\hat{p}},\underline{\hat{q}};T\right) =\frac{1}{2}%
\left\{ \left[ \hat{P}\left( \underline{\hat{p}}\right) +h\left(
\underline{\hat{y}},\underline{\hat{x}}\right) \right] ^{2}+\left( \frac{2\pi }{T}%
\right) \left[ \hat{Q}\left( \underline{\hat{q}}\right) \right] ^{2}\right\} ~.
\label{Htilde}
\end{equation}
entails an isochronous evolution, with arbitrary period $T$, of the center-of-mass $\hat{Q}\left( \underline{\hat{q}}\right) =\sum_{n=1}^{N}\hat{q}_{n}/N$, the total momentum $\hat{P}\left( \underline{\hat{p}}\right)
=\sum_{n=1}^{N}\hat{p}_{n}$ and all the 
coordinates
\begin{equation}
\underline{\tilde{x}}\left( t\right) =\underline{x}\left( \tau \left(
t\right) \right) ~,~~~\underline{\tilde{y}}\left( t\right) =\underline{y}%
\left( \tau \left( t\right) \right) ~,  \label{xytildet}
\end{equation}
where the "periodic time" $\tau(t)$ is given by
\begin{equation}
\tau \left( t\right) =A~\sin \left( \frac{2\pi \left(t-t_0\right)}{T}\right)\, ,
\label{taut}
\end{equation}%
with $A$ and $t_0 $  constant parameters that are functions of the initial values $Q\left( 0\right) $ and $P\left(
0\right) $, and of the conserved Hamiltonian $\tilde{H}\left( \underline{p},\underline{q};T\right) $.

We should emphasize that, having for any $\left\vert t\right\vert <<T,$%
\begin{equation}
\tau \left( t\right) =2\pi A\frac{ \left(t-t_0\right)}{T} +O\left[ \left( \frac{t-t_0}{T}\right) ^{2}%
\right],
\end{equation}%
the time evolution given by the isochronous Hamiltonian $\hat{H}$ approximates
that of $H$  up to a constant shift and rescaling of
time, hence up to corrections that remain small  as long as $t$ is in an assigned interval $[t_0-\tilde{T}/2,t_0+\tilde{T}/2]$, where $\tilde{T}<<T$, with $T$  arbitrary.
Moreover, different Hamiltonians $\hat{H}$ can be built corresponding to  different periodic changes of time and, if one requires that $\tau(t)\in\mathcal{C}^2$ in order to be compatible with  second order dynamical equations, $\tau(t)$ can be chosen to coincide with $t\in [t_0-\hat{T}/2,t_0+\hat{T}/2]$, while being a periodic function of $t$ with period $T>\hat{T}$.
In this case, the dynamics given by $\hat{H}$ will be identical to that of the system $H$ for any time  $t\in [t_0-\hat{T}/2,t_0+\hat{T}/2]$.

These results can be straightforwardly extended to general relativity.  Let us start with a metric $g_{\mu\nu}(y)$ given by

\begin{equation}
ds^{2}=g_{\mu \nu }(y)\,dy^{\mu }dy^{\nu }\,,  \label{metric y}
\end{equation}%
which is a solution of the Einstein's equations in vacuum \footnote{The extension  of the following arguments to the case of Einstein's equations coupled with fields is straightforward.}
\begin{equation}
R_{\mu \nu }(y)-\frac{1}{2}\, R(y)\,g_{\mu \nu }(y)=0\,,
\label{einstein equations y}
\end{equation}%
where $R_{\mu \nu }(y)$ and $R(y)$ are the
the Ricci and scalar curvature tensors constructed with the metric $g_{\mu \nu }(y)$. Let us consider the metric $\hat{g}_{\mu \nu }(x)$ in the
coordinate system $x$ defined by the complex periodic change of time
\be
y^0 = \tau(x^0)\,,\,\, \vec{y}= \vec{x}
\ee
where $\tau(x^0)=\tau(x^0+T)\in\mathcal{C}^3$ is a periodic function with period $T\in \mathbb{R}$, so that

\begin{equation}
\begin{array}{ll}
d\tilde{s}^{2}=\tau^\prime(x^{0})^{2}g_{00}\left( y\left( x\right) \right)
\,(dx^{0})^{2}+
\\
\\
2\,\tau^\prime(x^{0})\,g_{0k}\left( y\left( x\right) \right)
\,dx^{0}dx^{k}+g_{kh}\left( y\left( x\right) \right) \,dx^{k}dx^{h}\\
\\
\equiv \hat{g}_{\mu \nu }(x)\,dx^{\mu }dx^{\nu }\,, \label{metric x} &
\end{array}%
\end{equation}%
where $g_{\alpha \beta }\left( y\left( x\right) \right) =g_{\alpha \beta }\left( y^{0}=\tau (x^{0}),y^{k}=x^{k}\right) $. Indeed, $\hat{g}_{\mu \nu }(x)$ will be
a solution of  Einstein's equations in vacuum, which is degenerate at the hypersurfaces  $\tau^\prime(x^0)=0$. What is more, since $y^0(x^0)$ is a periodic function of $x^0$, the metric $\hat{g}_{\mu \nu }(x)$ is periodic in time.

Thus, considering a real periodic change of time $\tau(t)$, in principle one obtains periodic in time (in fact isochronous) but degenerate solutions of  Einstein's equations. However, remaining in the framework of general relativity, such spacetimes are physically unacceptable, since they are not locally Minkowskian. Indeed, we conclude that the equivalence principle forbids the use of periodic changes of time and the construction of isochronous universes.

However, if the class of admissible spacetimes is enlarged to include complex metrics, one has to enlarge the class of admissible diffeomorphysims to complex ones. Indeed, by means of a complex change of time, starting from any real solution $g_{\mu\nu}$ of  Einstein's equations, one can easily obtain nondegenerate but complex and periodic metrics $\hat{g}_{\mu\nu}$ that approximate $g_{\mu\nu}$ with arbitrary accuracy  over an arbitrarily long time interval.  For instance, choosing

\begin{figure}
	\begin{center}
		\includegraphics[height=2cm]{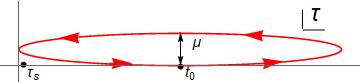}\qquad	 
	\end{center}
	\caption{We plot the complex change of time $\tau(t)$ in (\ref{complex change of time}) as a function of real $t$. If the spacetime $g_{\mu \nu}$ has a big-bang singularity at $y^0=\tau_s$, the periodic spacetime $\hat g_{\mu \nu}$ will be nonsingular, as $y^0(x^0)\neq \tau_s$ at any time $x^0$.}\label{figura2}	
\end{figure}

\be\label{complex change of time}
\tau(t)= t_0 + T \sin\left(\frac{t-t_0}{T}\right)+ i \mu \left[1- \cos\left(\frac{t-t_0}{T}\right)\right] 
\ee
for $t\in\mathbb{R}$, corresponding to the closed curve of the complex $\tau$ plane plotted in fig. \ref{figura2}, one has $\tau \simeq t$, and thus $y^0 \simeq x^0$ as long as   $|x-t_0|/T\ll 1$. Therefore, since $T$ is arbitrary,  the metrics $\hat{g}_{\mu\nu}$ and $g_{\mu\nu}$ are arbitrarily close over an arbitrarily long time interval, even though $\hat{g}_{\mu\nu}$ is periodic in the time $x^0$. Moreover, $\hat{g}_{\mu\nu}$ and $g_{\mu\nu}$ are diffeomorphic, indeed all the curvature invariants of $\hat{g}_{\mu\nu}$ can be calculated from those of $g_{\mu\nu}$.

Let see what happens to the geodesics of the two spacetimes. Let us suppose that the geodesics associated with the metric $g_{\mu\nu}$  are given by
\begin{equation}
Y =\left[ \lambda ,~\vec{Y}\left( \lambda \right) \right] ~,
\label{GeodesicFRW}
\end{equation}%
where the parameter $\lambda $ coincides with the time $y^{0} $ along the geodesics. It is easy to show (see \cite{isochronous spacetimes,isochronous cosmologies,isochronous Newtonian limit}) that the geodesics associated with the
periodic metric $\hat{g}_{\mu\nu}$ are given by the analogous formula
\begin{equation}
X = \left[
\mu ,~\vec{Y}\left( \tau \left( \mu \right) \right) \right]
\label{Geodesic isochronous 0}
\end{equation}%
and in this case the parameter $\mu $ coincides with the time $x^{0}=t$.
These geodesics are
open spiraling curves in spacetime, with the space coordinates evolving
periodically as functions of the time coordinate $x^{0}=\mu $.

It is worth  noting that, choosing the function $\tau(t)$ properly, one can avoid some of the singularities of the spacetime $g_{\mu\nu}$. For instance, if $g_{\mu\nu}$ is a FRW spacetime with a scale factor $a(y^0)$
\be\label{FRW}
ds^2  = (dy^0)^2-a(y^0)^2 d\vec{y}^2
\ee
which is a is solution of the Einstein's equations in the presence of a perfect fluid with equation of state $p = \omega \rho$, and with $p$ and $\rho$, the pressure and energy density of the fluid, one has

\begin{equation}
\begin{array}{ll}
a(y^0)= a_0\left[1+\sqrt{\frac{3\rho_0}{4}}\left(\omega+1\right)\left(y^0-y^0_0\right)\right]^{\frac{2}{3\left(\omega+1\right)}},
\\
\\
\rho = \rho_0 \left(\frac{a(0)}{a(y^0)}\right)^{-3\left(\omega+1\right)}\, .
\end{array} 
\end{equation}
The spacetime (\ref{FRW}) has a big-bang singularity at the time $y^0= y^0_0-2/\sqrt{3\rho_0}\left(\omega+1\right)\equiv \tau_s$, where $a(y^0)=0$. Imposing that the periodic change of time is such that $\tau(t)\neq \tau_s$ for any t, so that $y^0(x^0)\neq \tau_s$ for any time $x^0$, the periodic solution $\hat g_{\mu\nu}$ given by (\ref{metric x}) will evade the big-bang singularity. That means that the spacetime $\hat g_{\mu\nu}$ will be geodesically complete, so that the geodesics (\ref{Geodesic isochronous 0}) will be future and past extendible. Moreover, $\hat g_{\mu\nu}$ can be set to be arbitrarily close to $g_{\mu\nu}$ for  an arbitrarily large time interval, since corrections scale as $\sim T^{-1}$ and the period $T$ is arbitrary. In the case of the complex change of variables (\ref{complex change of time}), a non-singular periodic universe $\hat g_{\mu\nu}$ is obtained if $t_0\neq \tau_s$. Moreover, $\hat g_{\mu\nu}$ mimics the dynamics of $g_{\mu\nu}$ up to corrections $\sim |t-t_0|/T$ for any t such that $|t-t_0|/T\ll1$.

One might argue that, although arbitrarily close and connected by a (complex) diffeomorphism, the two evolutions enclosed in the spacetimes $g_{\mu\nu}$ and $\hat g_{\mu\nu}$ are physically inequivalent, as the time $y^0$ acquires a  complex part $\Im\left\{y^0\right\}$. This might be acceptable for infinitesimal $\Im\left\{y^0\right\}$. For instance, in QFT it is commonly accepted that the time variable has a small complex part, usually parametrized by the change of time $t\rightarrow(1-i\epsilon) t$ with $\epsilon$ infinitesimal, in order to ensure the convergence of path integrals. However, in general the complex component of $\tau(t)$ can be non-negligible, as in (\ref{complex change of time}) for large $\mu$. Moreover, one can object that the space components of the geodesics (\ref{Geodesic isochronous 0}) acquire a complex part as well, and it can be difficult to give a meaningful physical interpretation of this fact.

To circumvent these issues, we can thus look at periodic changes of time such that $\tau(t)\equiv t$ for $t$ in some time interval, say $t\in[-\hat{T}/2,\hat{T}/2]$. In this case, the two metrics $g_{\mu\nu}$ and $\hat{g}_{\mu\nu}$ will be identical over the arbitrary time interval $[-\hat{T}/2,\hat{T}/2]$, while $\hat{g}_{\mu\nu}$ will be periodic with arbitrary period. For instance,  one can consider the following complex function:

\be\label{complex time identical}
\tau(t)= \left\{\begin{array}{lr}
	t- n T  \qquad\text{for}   - \frac{\hat{T}}{2}+ n T \leq t \leq  \frac{\hat{T}}{2}+ n T\\
	Z(t-n T)\,\,\, \text{for}  \,\,\,  \frac{\hat{T}}{2}+ n T \leq t \leq - \frac{\hat{T}}{2}+ \left(n+1\right) T
\end{array} \right.
\ee
with $n$ integer, $t,T,\hat{T}\in \mathbb{R}$, $T>\hat{T}$, and $Z(t)\in \mathcal{C}^3$, as we want the periodic metric $\hat{g}_{\mu\nu}$ to be a solution of  second order Einstein's equations.

\begin{figure}
	\begin{center}
		\includegraphics[height=3cm]{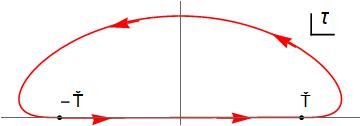}
	\end{center}
	\caption{We plot the complex change of time $\tau(t)$ in (\ref{complex time identical}) as a function of real $t$. One has $\tau(t)\equiv t$ for any $t\in[-\hat T, \hat T]$, while $\tau(t)$ is  periodic with period $T> 2\,\hat T$.  }\label{figura3}
\end{figure}

Since $\tau(t)$ is periodic, $Z(t)$ must be a complex function  $Z(t)=A(t)+iB(t)$ with with $A(t)$ and $B(t)$ real functions such that  $A(\hat{T})=\hat{T}$, $A(T-\hat{T})=-\hat{T}$,  $A^\prime(\hat{T})=A^\prime(T-\hat{T})=1$, $A^{\prime\prime}(\hat{T})=A^{\prime\prime}(T-\hat{T})=A^{\prime\prime\prime}(\hat{T})=A^{\prime\prime\prime}(T-\hat{T})=0$, and $B(\hat{T})=B(T-\hat{T})=B^\prime(\hat{T})=B^\prime(T-\hat{T})=B^{\prime\prime}(\hat{T})=B^{\prime\prime}(T-\hat{T})=B^{\prime\prime\prime}(\hat{T})=B^{\prime\prime\prime}(T-\hat{T})=0$ \footnote{An example of  functions $A(t)$ and $B(t)$ satisfying these conditions can be easily obtained assuming that  $A(t)$ and $B(t)$ are polynomials in $t$.}. The function $\tau(t)$ in (\ref{complex time identical}) is such that $\tau(t)\equiv t$ for $-\hat{T}/2<t<\hat{T}/2$, while it is periodic  with arbitrary period $T>\,\hat{T}$. An example of such a complex function $\tau(t)$ is plotted in fig. \ref{figura3}. Therefore, the metric $\hat g_{\mu \nu}$ will be  diffeomorphic to the metric $g_{\mu \nu}$, and it will be identical to $g_{\mu \nu}$ for $-\hat{T}<t<\hat{T}$. Indeed, $\hat g_{\mu \nu}$ and $g_{\mu \nu}$  yield an identical evolution over an arbitrary time interval  $t\in[-\hat{T},\hat{T}]$, where they are physically indistinguishable, while $\hat g_{\mu \nu}$ is periodic with a period $T$ that is also arbitrary.


\begin{figure}
	\begin{center}
		\includegraphics[height=1.5cm]{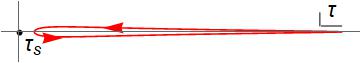}
	\end{center}
	\caption{Considering a FRW universe (\ref{FRW}) and  the complex change of time $\tau(t)$ in the plot, the corresponding periodic metric $\hat g_{\mu\nu}$ represents a nonsingular universe that contracts, has a bounce, and then expands forever, converging to (\ref{FRW}) at late times.  }\label{figura4}
\end{figure}

Finally, we note that, with a proper choice of $\tau(t)$, one can also construct  bouncing universes, and, in general, spacetimes with an inversion of the arrow of time. For instance, for the FRW metric (\ref{FRW}) and the function $\tau(t)$ plotted in Fig.\ref{figura4}, one has a contracting universe that reaches a minimum size corresponding to a minimum of $|a(\tau(x^0))|$, and then expands forever. As $\tau(t)\simeq t$ for large $t$, this universe converges to (\ref{FRW}) at late times.

The construction of  isochronous and bouncing spacetimes, which leads to the unpleasant implications described before, is based on the use of periodic diffeomorphisms, which are allowed if one considers complex metrics. In the case of real metrics, this construction was forbidden by the equivalence principle, which imposes that a physical acceptable spacetime must be locally Minkowskian, indeed nondegenerate.

In the framework of classical general relativity, there is no first principle capable of forbidding periodic complex diffeomorpisms, and the issues related to their use. However,  condition (\ref{criterion}), which has been introduced in \cite{Kontsevich}  following from the request that one can construct a physically meaningful quantum field theory over a complex spacetime, does the job. For instance, considering the FRW metric (\ref{FRW1}) and its complex periodic extension (\ref{flat metric complex 2}), the requirement (\ref{criterion}) implies that it must be $|Arg(\tau^\prime(t))|<\pi/2$ for physically acceptable spacetimes (\ref{flat metric complex 2}). This prevents the use of complex periodic changes of time and complex time inversions, since for these time transformations one necessarily has $\Re\{\tau^\prime(t)\}=0$ at some time. This argument, can be extended to more general complex isochronous spacetimes as those in Eq. (\ref{metric x}), so that one concludes that (\ref{criterion}) excludes complex periodic diffeomorphisms from the class of  physically
acceptable changes of coordinates.

In conclusion, we have shown that, if complex metrics are considered, for any real spacetime $g_{\mu\nu}$, by means of complex changes of time, it is possible to build infinitely many periodic or bouncing nondegenerate complex metrics 
$\hat g_{\mu\nu}$, which are indistinguishable from $\hat g_{\mu\nu}$ over an arbitrary long time interval. This result leads to a physically unacceptable arbitrariness, which might be seen as an inconsistency of the theory. In the case of general relativity, the equivalence principle forbids the use of real periodic changes of time, while it does not exclude the use of periodic diffeomorphisms if one considers complex metrics. However, the condition (\ref{criterion}) forbids complex periodic changes of time and complex time inversions, preventing  the construction of isochronous and bouncing spacetimes, and avoiding all the related issues. For this reason,  (\ref{criterion})
can be viewed as a quantum-gravity generalization of the equivalence principle.

\section*{Acknowledgements} The author wishes to thank P. Bomans, F. Calogero, and E. Witten for several useful discussions on the draft version of this manuscript.


\begin{thebibliography}{99}








\bibitem{calogero1} F. Calogero and F. Leyvraz,
J. Phys. A.: Math. Theor. \textbf{40},
12931-12944 (2007).

\bibitem{calogero2} F. Calogero and F. Leyvraz, 
J. Nonlinear Math. Phys. \textbf{16}, 311-338 (2009).

\bibitem{calogero3}F. Calogero and F. Leyvraz, 
Lett. Math. Phys. \textbf{96}, 37-52
(2011). 

\bibitem{calogero4} F. Calogero, 
Soc. \textbf{A 369}, 1118-1136 (2011). 




\bibitem{calogero6} F. Calogero and F. Leyvraz, 
J. Nonlinear Math. Phys. \textbf{16}, 311-338
(2009). 

\bibitem{calogero7}F. Calogero and F. Leyvraz, 
Lett. Math. Phys. \textbf{96}, 37-52 (2011). 

\bibitem{calogero8}F. Calogero,
Phil. Trans. R.
Soc. \textbf{A 369}, 1118-1136 (2011).

\bibitem{calogero book}F. Calogero, \textit{Isochronous
	systems}, Oxford University Press, 2008 (marginally updated paperback
edition, 2012).




\bibitem{isochronous cosmologies} F. Briscese and F. Calogero, 
Int. J. Geom. Meth. Mod. Phys. \textbf{11},
1450054 (2014). 	arXiv:1402.0704 [gr-qc].

\bibitem{isochronous spacetimes} F. Briscese and F. Calogero, 
Acta Appl. Math. \textbf{137} (2015) 3-16. arXiv:1406.7156 [gr-qc].

\bibitem{isochronous Newtonian limit} F. Briscese and F. Calogero,  Int. J. Geom. Meth. Mod. Phys. {\bf 15} (2018) 1850101. arXiv:1803.01105 [gr-qc].

J. Phys. Conf. Ser. \textbf{626}
(2015) no.1, 012004.









\bibitem{Gibbson} G. W. Gibbons, S. W. Hawking, 
Phys. Rev. D {\bf 15} (1977) 2752-6.


\bibitem{Gibbson 2} G. W. Gibbons, S. W. Hawking, M. J. Perry, 
Nucl. Phys. B {\bf 138} (1978) 141-50.

\bibitem{Haliwell} J. J. Haliwell, J. B. Hartle, 
Phys. Rev. D {\bf 41} (1990) 1815-34.

\bibitem{Hartle} J. B. Hartle, S. W. Hawking, 
Phys. Rev. D {\bf 28} (1983) 2960-75.

\bibitem{Louko} J. Louko, R. Sorkin, 
Class. Quant. Grav. {\bf 14} (1997) 179-204, arXiv:gr-qc/9511023.



\bibitem{Kontsevich} M. Kontsevich and G. B. Segal, 
Quart. J. Math. {\bf 72} (2021) 673-99, arXiv::2105.10161.


\bibitem{Witten} E. Witten, arXiv:2111.06514 [hep-th].

\bibitem{a1} G. J. Loges, G. Shiu, N. Sudhir,
e-Print: 2203.01956 [hep-th].


\bibitem{a2} Y. Chen,
e-Print: 2202.04741 [hep-th].


\bibitem{a3} B. Muhlmann,
e-Print: 2202.04549 [hep-th].

\bibitem{a4} U Moitra, S. Kumar Sake, S. P. Trivedi,
e-Print: 2202.03130 [hep-th].

\bibitem{a5} J.-M. Schlenker, E. Witten, 
e-Print: 2202.01372 [hep-th].

\bibitem{a6} N. Turok, L. Boyle,
e-Print: 2201.07279 [hep-th].

\bibitem{a7} S. K. Asante, B. Dittrich, J. Padua-Argüelles,
e-Print: 2112.15387 [gr-qc].

\bibitem{a8} M. David, V. Godet, Z. Liu, L. A. Pando Zayas,
e-Print: 2112.09444 [hep-th].

\bibitem{a9} C. Jonas, J.-L. Lehners, V. Meyer, Phys.Rev.D {\bf 105} (2022) 4, 043529 • e-Print: 2112.07986 [hep-th].

\bibitem{a10} K. Rajeev,
e-Print: 2112.04522 [gr-qc].

\bibitem{a11} E. Andriolo, T. Orchard, C. Papageorgakis, e-Print: 2112.00040 [hep-th].

\bibitem{a12} A. Cabo-Bizet,
e-Print: 2111.14942 [hep-th].

\bibitem{a13} C. Peng, J. Tian, J. Yu, 
e-Print: 2111.14856 [hep-th].

\bibitem{a14} M. Visser,
e-Print: 2111.14016 [gr-qc].

\bibitem{a15} M. Medevielle, T. Mohaupt, G. Pope, JHEP 02 (2022) {\bf 048}, e-Print: 2111.09017 [hep-th].

\bibitem{a16} J. L. Lehners,  Phys. Rev. D {\bf 105} (2022) 2, 026022 • e-Print: 2111.07816 [hep-th].

\bibitem{a17} S. Bondarenko,
e-Print: 2111.06095 [gr-qc].

\bibitem{a18} P.  J. Martinez, G. A. Silva, JHEP {\bf 03} (2022) 003, e-Print: 2110.07555 [hep-th].

\bibitem{a19} V. V. Belokurov, E. T. Shavgulidze, JHEP 02 (2022){ \bf 112}, e-Print: 2110.06041 [hep-th].

\bibitem{a20} P. Mao, Weicheng Zhao,
 JHEP 01 (2022) {\bf 030 }, e-Print: 2109.09676 [gr-qc].









\end{thebibliography}
\end{document}